\documentclass[a4paper,fleqn,usenatbib]{mnras}

\usepackage{newtxtext,newtxmath}
\usepackage[T1]{fontenc}
\usepackage{ae,aecompl}

\usepackage{longtable}
\usepackage{graphicx}	% Including figure files
\usepackage{amsmath}	% Advanced maths commands
\usepackage{amssymb}	% Extra maths symbols

\usepackage{pdflscape}
\usepackage{multicol}

\title[ WRLOF - a path of very wide SySt to SN Ia?]{Wind Roche Lobe Overflow as a way to make type Ia supernova from the widest symbiotic systems}
\author[I{\l}kiewicz et al.]{Krystian I{\l}kiewicz,$^{1}$\thanks{E-mail: ilkiewicz@camk.edu.pl} Joanna Miko{\l}ajewska,$^{1}$ Krzysztof Belczy{\'n}ski$^{1}$, \newauthor Grzegorz Wiktorowicz$^{2,3}$ and Paulina Karczmarek$^{4}$ \\
$^{1}$Nicolaus Copernicus Astronomical Centre, Bartycka 18, 00716 Warsaw, Poland \\
$^{2}$National Astronomical Observatories, Chinese Academy of Sciences, Beijing 100012, China\\
$^{3}$School of Astronomy \& Space Science, University of the Chinese Academy of Sciences, Beijing 100012, China\\
$^{4}$Warsaw University Observatory, Al. Ujazdowskie 4, 00-478 Warszawa, Poland }

\date{Accepted XXX. Received YYY; in original form ZZZ}

\pubyear{2015}

\begin{document}
\label{firstpage}
\pagerange{\pageref{firstpage}--\pageref{lastpage}}
\maketitle

% Abstract of the paper
\begin{abstract}
Symbiotic stars  are interacting binaries with one of the longest orbital periods. Since they can contain a massive white dwarf with a high accretion rate they are considered a promising type Ia supernovae (SNe Ia) progenitors. Among symbiotic binaries there are systems containing a Mira donor, which can have orbital periods of a few tens of years and more. This subclass of symbiotic stars due to their very large separation usually was not considered promising SNe Ia progenitors. We analysed evolution of one of the well studied symbiotic star with a Mira donor, V407 Cyg.  We showed that the standard evolution model predicts that the system will not become a SN Ia. However, by simply adding a Wind Roche Lobe Overflow as one of the mass transfer modes we predict that the white dwarf in V407~Cyg will reach the Chandrasekhar limit in 40--200 Myr. %This finding is particularly interesting to the study of SN Ia environments.
\end{abstract}

% Select between one and six entries from the list of approved keywords.
% Don't make up new ones.


%%%%%%%%%%%%%%%%%%%%%%%%%%%%%%%%%%%%%%%%%%%%%%%%%%

%%%%%%%%%%%%%%%%% BODY OF PAPER %%%%%%%%%%%%%%%%%%

\section{Introduction}

Type Ia supernova (SN Ia) is a thermonuclear outburst of a carbon--oxygen (CO) white dwarf (WD). Many scenarios have been proposed for progenitors of SNe Ia, which include a single-degenerate (SD) scenario, where WD is accreting mass from a companion until it reaches a Chandrasekhar mass (M$_{\mathrm{CH}}$) and as a result explodes as a SN Ia \citep[e.g.][]{1973ApJ...186.1007W,2018RAA....18...49W}. A double-degenerate (DD) scenario involves a merger of two WDs \citep{1984ApJS...54..335I,1984ApJ...277..355W}. Similarly, in the core-degenerate (CD) scenario the SN~Ia outburst is due to a common-envelope phase of a binary with a WD and an asymptotic giant branch (AGB) star, which leads to a merger of the WD and a degenerate core of the AGB star \citep[e.g.][]{2013IAUS..281...72S,2017MNRAS.464.3965W}. There are also scenarios involving SN~Ia on a sub-M$_{\mathrm{CH}}$ WD. In one of the scenarios the SN~Ia is triggered by a helium shell detonation \citep{2017Natur.550...80J}. It is also speculated that SN~Ia can also be produced by a single, isolated WD when it transfers from a liquid to the solid state \citep{2015MNRAS.448.2100C}. Unfortunately, there are no observations of systems that produced SN Ia before the explosion, hence none of the scenarios is directly confirmed trough observations. Moreover, each of the scenarios has its problems and most probably more than one scenario is responsible for the observed SN~Ia population. Possible progenitors of SN~Ia, together with their advantages and disadvantages are presented e.g. in \citet{2014ARA&A..52..107M}, \citet{2018PhR...736....1L} and \citet{2018arXiv180503207P}.

Symbiotic stars (SySt) are long-period interacting binaries with a red giant (RG) as a mass donor. Most commonly a white dwarf is an accretor, but in some systems a neutron star (NS) is observed  (see \citealt{2012BaltA..21....5M} for a recent review of SySt). Since some of them contain a massive WD with a high accretion rate, they have had been proposed as possible SN Ia progenitors (see e.g. \citealt{1999ApJ...522..487H}; \citealt{2009MNRAS.396.1086L};  \citealt{2016MNRAS.457..822B}; \citealt{2017ApJ...847...99M}, \citealt{2017arXiv171003965L}). SySt can be considered as SN~Ia progenitors both in the SD scenario, as well as in the DD scenario, in which case the RG evolves into WD, leading to a WD+WD binary. \citet{2012Sci...337..942D} showed that the evolution of the SN Ia PTF 11kx outburst was consistent with a symbiotic nova progenitor, although other scenarios for this outburst have been proposed as well \citep{2013MNRAS.431.1541S}. A review of most promising SN Ia progenitors among SySt is presented in \citet{2013IAUS..281..162M}.

Kepler`s Supernova Remnant,  a remnant of SN Ia, has dust features consistent with a SD scenario and a massive  AGB donor \citep{2012ApJ...755....3W}.  It's progenitor could had been very similar to the SySt V407~Cyg \citep{2012A&A...537A.139C,2013IAUS..281..162M}. V407~Cyg is a SySt containing a massive Mira donor on  AGB  \citep{2003AstL...29..405T,2003MNRAS.344.1233T} and a massive WD \citep[e.g.][]{2012BaltA..21...68H} -- a system almost identical to the speculated progenitor of Kepler's supernova. The presence of the Mira providing a vast amount of dust into the system V407~Cyg implies that it belongs to the D--type ($dust$) SySt, which have the widest orbits among SySt with expected orbital periods of the order of a few tens of years and more \citep[e.g.][]{2009AcA....59..169G}.

In this study we use population synthesis models to predict the fate of V407~Cyg and its viability as a SN Ia progenitor in the frame of the SD scenario. In particular, we study importance of including improved formulae for mass transfer trough wind accretion in context of such wide binary systems as V407~Cyg, which rarely can have high mass accretion rates in the standard picture of evolution. The model parameters are presented in section~\ref{model_sec}. The results are presented in section~\ref{results_sec} and discussed in section~\ref{discussion_sec}. Summary of our study follows in section~\ref{summary_sec}.

\section{Model}\label{model_sec}

\subsection{Adopted binary parameters}\label{binpar_sec}
The binary period of V407 Cyg is very uncertain, although we expect that it must be of a few tens of years or longer.  Here, we adopt  the orbital period of 43 years suggested by \citet{1990MNRAS.242..653M} based on long-term changes in Mira brightness, which is consistent with a lower limit of the orbital period derived by \mbox{\citeauthor{2011A&A...527A..98S}} \citeyearpar{2011A&A...527A..98S}. We note that larger values of the orbital period may affect our estimates.  Since there is no published information about possible eccentricity of the system we assumed a circular orbit.
 
There are many observational evidences that the WD in V407~Cyg is very massive \citep[e.g.][]{2012ApJ...748...43N}. Some studies suggest that the WD is as massive as 1.35--1.37~M$_\odot$ \citep{2012BaltA..21...68H}. The fact that the evolution of the recent nova outburst was very similar to the outburst of a symbiotic recurrent nova RS~Oph hints that the mass of a WD is at the very least $\gtrsim1.2$~M$_\odot$ \citep{2010arXiv1011.5657M}. In our parameter space we explore the masses of WD M$_{\mathrm{WD}}>1$~M$_\odot$, although we point out that the mass of WD is most probably significantly larger than 1~M$_\odot$. 

Since carbon-oxygen white dwarfs (CO~WDs) are not born with masses exceeding 1.1~M$_\odot$, unless the WD did grow mass in the prior evolution of the system, an oxygen-neon white dwarf (ONe WD) is more likely to be present in V407~Cyg. While there is no evidence of CO~WD or ONe~WD in V407~Cyg, in principle growing a CO~WD to such mass is possible. This is the case in another SySt, RS~Oph, where there is a CO~WD with mass $\ge$1.2~M$_\odot$ \citep{2017ApJ...847...99M}. The type of WD present in the system will not significantly change its evolution due to the fact that with the same accretion rate and the same WD mass, both CO~WD and ONe~WD have the same accumulation rates of the mass on their surface in our code. The only difference in the evolution of the system would be when ONe~WD will exceed the Chandrasekhar limit it will transform into a NS in a accretion-induced collapse (AIC). In the case of CO~WD, after exceed the Chandrasekhar limit a SN~Ia will occur. Since we study V407~Cyg in the SN~Ia context we calculated models with a CO~WD. For comparison we also calculated a set of models with a ONe~WD. 

The RG in  V407~Cyg is Li--rich (\citealt{2003MNRAS.344.1233T}; \citealt{2003AstL...29..405T}). 
The Li-enrichment in V407 Cyg and in other RG variables with similarly long periods is most likely due to hot bottom burning (HBB; e.g. \mbox{\citealt{1999IAUS..191...91G}}), i.e. nucleosynthesis at the bottom of the outer convective zone in a massive RG star. The occurrence of HBB  is limited to relative narrow range of masses, 4--8~M$_\odot$ and bolometric luminosity from $-6$~mag to $-7$~mag \citep{1990ApJ...361L..69S}, which we adopted in our analysis.

The effective temperature of the Mira is poorly defined, since it is changing during a pulsation period. The spectral type of the RG in V407~Cyg at different pulsation phases was derived by \citet{1990MNRAS.242..653M} to be M6~III and by \mbox{\citet{2013ApJ...770...28H}} to be M7~III. This corresponds to effective temperatures T$_{\mathrm{eff}}$=2900--3100~K. Since the temperature is poorly defined we adopted in our model  T$_{\mathrm{eff}}$=2700--3300~K. 

\subsection{Population synthesis code}
In our study we employed the StarTrack population synthesis code \citep{2002ApJ...572..407B,2008ApJS..174..223B}. The code included mass transfer trough the standard \citet{1944MNRAS.104..273B} wind accretion and trough Roche lobe overflow (RLOF). We expanded the code by adding a $Wind$ Roche lobe overflow (WRLOF; \citealt{2012BaltA..21...88M}). In order to implement WRLOF we adopted the prescription in eq.~9 of \citet{2013A&A...552A..26A}:
\begin{equation}
\beta_{\mathrm{acc}}=\min \left \{ \frac{25}{9}\mathrm{q}^2 \left[-0.284\left(\frac{\mathrm{R}_\mathrm{d}}{\mathrm{R}_\mathrm{L}}\right)^2+0.918\frac{\mathrm{R}_\mathrm{d}}{\mathrm{R}_\mathrm{L}}-0.234 \right] \textrm{ , } 0.5 \right \}
\end{equation}
where $\beta_{\mathrm{acc}}$ is the ratio between the mass accreted by the secondary star and the mass lost by donor star per unit of time, q$=\mathrm{M}_\mathrm{secondary}/\mathrm{M}_\mathrm{donor}$, $\mathrm{R}_\mathrm{d}$ is the dust formation radius and $\mathrm{R}_\mathrm{L}$ is Roche-lobe radius of the mass donor. The dust formation radius can be calculated with eq. 4 of \citet{2007ASPC..378..145H}:
\begin{equation}
\frac{\mathrm{R}_\mathrm{d}}{\mathrm{R}_\mathrm{*}}=\frac{1}{2}\left( \frac{\mathrm{T}_\mathrm{d}}{\mathrm{T}_\mathrm{eff}}\right)^{-\frac{4+p}{2}}
\end{equation}
where $\mathrm{R}_\mathrm{*}$ is the mass donor radius, $\mathrm{T}_\mathrm{d}$  is the dust condensation temperature and $p$ is a parameter characterising wavelength dependence of the dust opacity. We assumed dust consisting of  amorphous carbon grains, for which $\mathrm{T}_\mathrm{d}\simeq1500$K and $p\simeq 1$ \citep{2007ASPC..378..145H}. In our model we calculated mass accretion rate at each time step using the WRLOF and the standard \citet{1944MNRAS.104..273B} accretion, and used the larger value from these two as the mass accretion rate.

Typically, when a WD is accreting matter recurrent nova outbursts are expected. During a nova outburst large amount of matter is ejected from the WD. It remains controversial whether the WD is accreting more mass during a nova cycle than it losses during outburst, i.e. how much (if any) of the accreted mass is retained. In principle, the theoretical and observational considerations  show that at least in some cases the WD mass can grow (\citealt{2005ApJ...623..398Y}; \citealt{2007ApJ...663.1269N}; \citealt{2017ApJ...847...99M}). In our study, in the case of CO~WD we used mass accumulation formulae from \citet{2004ApJ...601.1058I}. Namely, in the case of mass accretion rate lower than $10^{-11}$~M$_\odot$yr$^{-1}$ all accreted mass is ejected in nova explosions, in the case of mass accretion rate between $10^{-11}$ and $10^{-6}$~M$_\odot$yr$^{-1}$ the fraction of accreted mass retained on the WD is interpolated from \citet{1995ApJ...445..789P} models, and in the case of mass accretion rate grater than $10^{-6}$~M$_\odot$yr$^{-1}$ all of the acretted mass is retained on the WD. Moreover, we adopted M$_{\mathrm{CH}}$=1.40~M$_\odot$ \mbox{\citep[e.g.][]{2015MNRAS.446.1924H}}. In the case of ONe~WD we used the same accumulation formulae as in the case of CO~WD \citep{2008ApJS..174..223B}.

In order to study V407~Cyg we evolved a single star from the Main Sequence (MS) to the RG stage. We stopped the single star evolution when the star met the parameters of the RG star adopted in our model, i.e. the mass, effective temperature and luminosity (section~\ref{binpar_sec}). Subsequently, we modelled a binary with a RG with the same structure as the one obtained from the single star evolution. The second star in the binary was a WD on a circular orbit consistent with the accepted orbital period. In order to fully explore the adopted parameter space we chose a range of masses on the MS by trial and error until we found the range for which the RG could evolve to the desired mass, luminosity and effective temperature. Using this procedure we were able to model systems with the RG mass in a range 4.00--7.30~M$_\odot$. Stars with masses higher than 7.30~M$_\odot$ did not reach the desired luminosity and effective temperature in a single star evolution models. We carried out calculation in a grid of models covering 300 stars evenly spaced in the zero age MS mass space. The second part of the grid consisted of 100 WDs evenly spaced in the current mass ranging from 1.0~M$_\odot$ up to 1.399~M$_\odot$. The prior evolution, leading to the SySt phase is beyond the scope of this paper. Our method is similar to that adopted by \citet{2011ApJ...732...70L} to study X-ray binaries evolution.

\section{Results}\label{results_sec}
\begin{figure}\centering
\resizebox{\hsize}{!}{\includegraphics{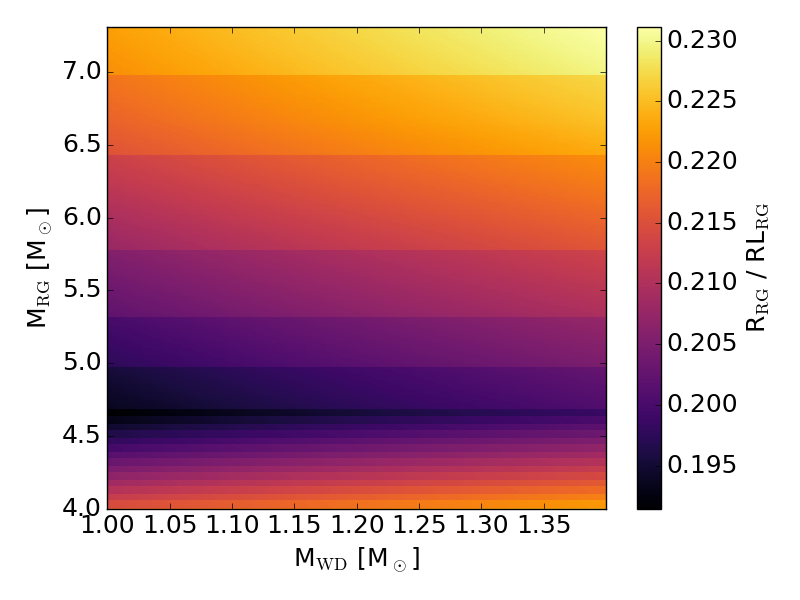}}
  \caption{The Roche-lobe filling factor of the RG in V407~Cyg-like system, depending on component masses. Orbital period was assumed to be $P=43$ years.}
\label{filling_rat}
\end{figure}

The  Roche-lobe filling factor of the RG in V407~Cyg, defined as the ratio of the stellar radius to the volume Roche-lobe radius, is presented in Fig.~\ref{filling_rat}. This ratio is small in all our models, exactly as expected for a D--type SySt. This means that no other channel proposed to enhance mass accretion rate in SySt, such as tidally enhanced RG wind \citep[e.g.][]{2009MNRAS.396.1086L}, is applicable in the case of V407~Cyg.

\begin{figure*}\centering
\resizebox{0.49\hsize}{!}{\includegraphics{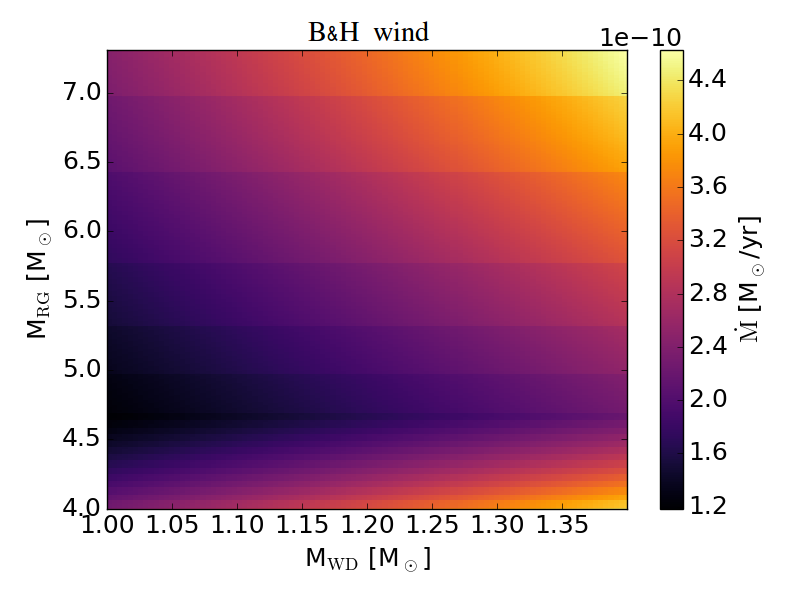}}
\resizebox{0.49\hsize}{!}{\includegraphics{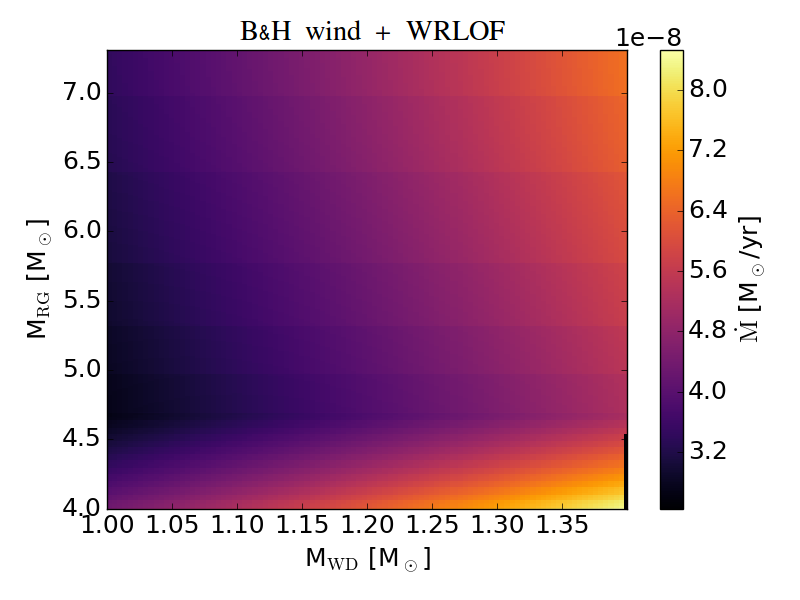}}
  \caption{Current mass accretion rate in V407~Cyg. Left panel: accretion only trough the standard \citet{1944MNRAS.104..273B} wind accretion. Right panel: Accretion trough \citet{1944MNRAS.104..273B} process and Wind-RLOF.   }
\label{acc_rate}
\end{figure*}

In Fig.~\ref{acc_rate} the calculated mass accretion rate is presented. Using only standard wind accretion (left panel) the mass accretion rate is over two orders of magnitude lower than when WRLOF is included (right panel). This indicates that WRLOF can be important mass transfer mode for systems such as V407~Cyg. Moreover, since WRLOF is based on more realistic estimates than the simple estimates of the standard wind accretion \citep{2012BaltA..21...88M} its inclusion enables to perform more realistic predictions about evolution of such systems.

\begin{figure*}\centering
\resizebox{\hsize}{!}{\includegraphics{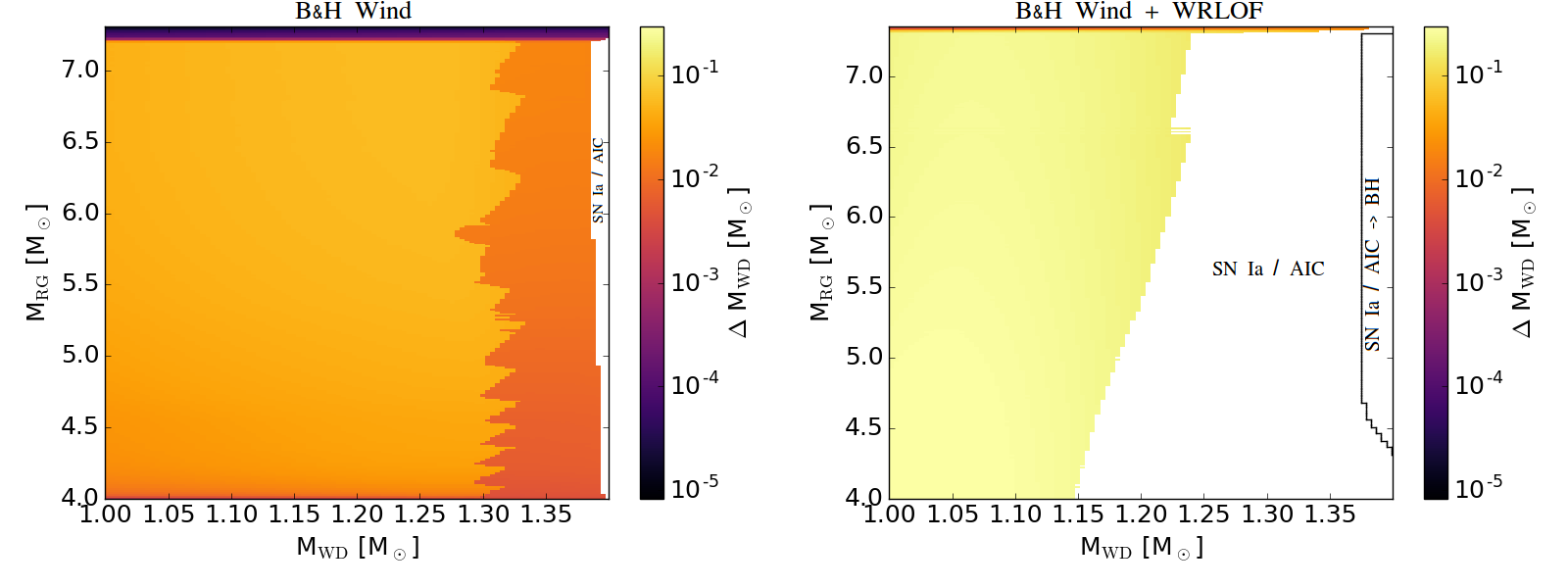}}
  \caption{Amount of accumulated mass on the WD during future evolution of system as a function of current system parameters. The white space denotes the models in which the WD reaches SN Ia in the case of CO~WD and AIC in the case of ONe WD. Left panel: accretion only trough the standard \citet{1944MNRAS.104..273B} wind accretion. Right panel: Accretion trough \citet{1944MNRAS.104..273B} process and Wind-RLOF. The AIC -> BH region denotes a parameter space where the compact object in the system accumulated enough mass to transform a NS to BH after the AIC.  }
\label{mass_acc}
\end{figure*}

We calculated the amount of mass that is accumulated onto a WD before the RG evolves into a degenerate object (Fig.~\ref{mass_acc}). In the models where only the standard \citet{1944MNRAS.104..273B} wind accretion is included the WD has to have essentially a mass equal to the Chandrasekhar limit in order to explode as SN~Ia or experience AIC. When we consider models with WD masses larger than 1.2~M$_\odot$ \citep{2010arXiv1011.5657M} only 7\% of the models, that have the most massive WDs, reach M$_{\mathrm{CH}}$. Moreover, none of the WDs with masses in range M$_{\mathrm{WD}}=1.35-1.37$~M$_\odot$ \citep{2012BaltA..21...68H}, estimated from the modelling of the evolution of the nova outburst, are massive enough to reach M$_{\mathrm{CH}}$. On the contrary, when the WRLOF is included in modelling, as much as 90\% and 97\% of the models exhibited SN~Ia or AIC for the 1.2~M$_\odot$ lower limit \citep{2010arXiv1011.5657M} and the M$_{\mathrm{WD}}=1.35-1.37$~M$_\odot$ estimate \citep{2012BaltA..21...68H} respectively.  Moreover, even models with the least massive RGs systems with WD masses as low as 1.15~M$_\odot$ could exhibit SN~Ia or AIC. This shows that WRLOF dramatically changes the predicted system evolution and including this mass transfer mode to population synthesis of the D--type SySt is crucial for estimating their plausibility as SN~Ia or AIC system progenitors. The WD in this scenario reached the Chandrasekhar limit in 40--200 Myr. In the case of most massive WDs and evolution including WRLOF, only the models with most massive RGs did not exhibit SN~Ia outburst or AIC. This is because such a massive RG, after reaching the temperature and luminosity of the RG in V407~Cyg, evolves towards core collapse, supernova II and NS formation, aborting mass accumulation onto WD and ending the system evolution as a WD+NS binary. 

Prolonged RLOF accretion onto WD in close binary system may lead to the formation of a low mass black hole. If accretion induced collapse of WD to NS does not disrupt or signifcantly widens the binary orbit, the RLOF may continue and provided that the donor star has enough mass NS may eventually collapse to a BH (e.g., fig.~2 of \citealt{2004ApJ...603..690B}). This process will naturally produce low-mass BHs that may potentially fill the apparent mass gap between known NSs and BHs (the lack of compact objects in mass range $2-5$~M$_\odot$; e.g. \citealt{1998ApJ...499..367B}, \citealt{2010ApJ...725.1918O}, \citealt{2012ApJ...757...91B}, \citealt{2016MNRAS.458.3012W}). In our models we assumed that the NS will collapse to a BH at M$_\mathrm{NS}$=2~M$_\odot$ \citep{2004ApJ...603..690B}. The BH in fact was created in the case of 10\% models of our models with ONe WDs with masses larger than 1.2~M$_\odot$ \citep{2010arXiv1011.5657M} an in none of the models with WD masses in range M$_{\mathrm{WD}}=1.35-1.37$~M$_\odot$ \citep{2012BaltA..21...68H}. In the models in which the WD experienced AIC and then the resulting NS collapsed into a BH the donor ends its evolution as a WD, which means that V407~Cyg can be considered as a potential progenitor of a WD+BH binary.

For consistency check, we calculated expected nova shell mass. For this calculations we assumed M$_{\mathrm{WD}}$=1.2~M$_\odot$ and mass accretion rate $\dot{\mathrm{M}}$=5$\times10^{-8}$~M$_\odot$/yr (Fig.~\ref{acc_rate}). From fig.~4~in~\citet{2016ApJ...819..168H} one can see that the expected nova cycle duration for assumed parameters is D$\simeq$100~yr. From fig.~1~in~\citet{2016ApJ...819..168H} we see that in this case the effective rate, at which the mass is accumulated onto the WD (i.e. mass that is retained on the WD during a nova cycle divided by a nova cycle duration), is of order of $\dot{\mathrm{M}}_{\mathrm{eff}}\simeq1\times10^{-8}$~M$_\odot$/yr. The expected nova shell mass is then D$\times(\dot{\mathrm{M}}-\dot{\mathrm{M}}_{\mathrm{eff}})\simeq4\times10^{-6}$~M$_\odot$. This is consistent with the shell mass of $\sim10^{-6}$~M$_\odot$ estimated from observations (\citealt{2010Sci...329..817A}; \citealt{2012ApJ...761..173C}).

\section{Discussion}\label{discussion_sec}

We point out that the scenario in which WRLOF leads to SN~Ia is different than the one proposed by \citet{2009MNRAS.396.1086L}, \citet{2011ApJ...735L..31C} and \citet{2017arXiv171003965L}. In these works the authors proposed changes in mass-loss from the RG. This is different from the WRLOF, since in order to change mass loss from the RG a  strong interaction between a WD and RG is needed. In the case of WRLOF, the stellar wind is originally unaffected by the WD, but then it is funnelled onto WD when the wind itself fills the Roche lobe of the RG. Hence, tidal forces are not needed and the model applies to much wider systems.

The caveat of our study is that \citet{2012ApJ...761..173C} argued  that the environment of V407~Cyg is not typical of SNe Ia. On the other hand, more recently \citet{2014MNRAS.443.1370D} showed that the observational constrains on the environment of SN Ia may not apply to systems in which there were classical nova outburst in the past (one of which was observed in V407~Cyg). Moreover, \citet{2014MNRAS.443.1370D} showed that the circumstellar medium shaped by nova explosions seem to better reproduce observational features of some SN Ia. We additionally point out  that the environment shaped by WRLOF is more complex \citep{2012BaltA..21...88M} then the one in the model used by \citet{2012ApJ...761..173C}. %Moreover, \citet{2012ApJ...761..173C} in their model used a RG mass of (...), as suggested by \citet{1990MNRAS.242..653M}. This mass is inconsistent with the never limits of the RG mass (see section \ref{binpar_sec}). 

Thus far the main arguments against SySt as promising SN~Ia progenitors from existing population synthesis models of SySt (\citealt{1995ApJ...447..656Y}; \citealt{1996ApJ...466..890Y}; \citealt{2006MNRAS.372.1389L}; \citealt{2010AstL...36..780Y}) were that: (i) the typical mass of the accretor is small ($\sim$0.5~M$_\odot$), (ii) mass accumulation of matter transferred to the WD is small \citep[see e.g.][]{1990LNP...369..390K} and (iii) they often evolve towards dynamically unstable mass transfer and formation of a common envelope. We point out that there are at least a few observed SySt with massive WDs, including four symbiotic recurrent novae \citep{2013IAUS..281..162M,2017ApJ...847...99M}. Moreover, population synthesis models of SySt are not reproducing the observed distribution of orbital periods \citep{2012BaltA..21....5M}, which hints that theoretical predictions about the SySt population might be yet inaccurate. In the case of mass accumulation onto the WD, the theoretical predictions are an active field of research, but it seems that the WD can accumulate mass as indicated by both theory \citep{2016ApJ...819..168H} and observations \citep{2017ApJ...847...99M}. Last but not least, we point out that in such systems as V407~Cyg the binary components separation is too big to reach a common envelope phase. Concluding, our simulations show that, contrary to previous suggestions based mostly on population synthesis models, binary systems like V407~Cyg could be promising SN~Ia progenitors.

\section{Summary}\label{summary_sec}
In this work we analysed predicted evolution of a D--type SySt V407~Cyg. Our analysis showed that in the framework of Chandrasekhar mass SN~Ia explosions, the standard picture may need an update for wide symbiotic systems. In the classical form of wind accretion implementation, the WD would need to have almost exactly Chandrasekhar mass in the present in order to reach a SN~Ia. On the other hand, when WRLOF is included as one of the modes of mass transfer, there is 90\% probability of encountering a SN~Ia when most conservative estimates of the WD mass and a CO~WD are adopted. The probability reached 97\% for the most accurate WD mass estimate (M$_{\mathrm{WD}}=1.35-1.37$~M$_\odot$; \citealt{2012BaltA..21...68H}). This shows that WRLOF is important factor in studying the D--type SySt evolution. Moreover, WRLOF could be a vital element in evolution of SySt with the longest orbital periods in context of SN~Ia. Given the problems of SD scenario \citep{2018PhR...736....1L} SySt can alternatively be considered as progenitors of peculiar SN~Ia \citep[see e.g.][]{2017hsn..book..317T}.

Since the distribution of circumbinary material shaped by WRLOF is more complicated than in the simpler models, such as e.g. \citet{1944MNRAS.104..273B} model, the SN~Ia produced in the proposed scenario could be interesting in the study of environments of SN~Ia. In particular, the large separation of considered binary and complex morphology of circumbinary material could influence lightcurve and spectral evolution of a SN~Ia outburst.

We also showed that if the WD in V407~Cyg is a ONe WD and not a CO~WD, the WD in the system will probably experience AIC. In the models with the highest initial WD masses the NS after AIC can accumulate enough mass in order to collapse into a BH, which means that V407~Cyg can be considered a potential a progenitor of a WD+BH binary with a low mass BH in the apparent mass gap between
known NSs and BHs.

\section*{Acknowledgements}

KI has been financed by the Polish Ministry of Science and Higher Education Diamond Grant Programme via grant 0136/DIA/2014/43 and by the Foundation for Polish Science (FNP) within the START program. This study has been supported in part by the Polish National Science Center (NCN) grants OPUS 2017/27/B/ST9/01940  and MAESTRO 2015/18/A/ST9/00746. KB also acknowledges support from the  NCN grants: Sonata Bis 2 2012/07/E/ST9/01360, and OPUS 2015/19/B/ST9/01099 and 2015/19/B/ST9/03188. GW is partly supported by the President's International Fellowship Initiative (PIFI) of the Chinese Academy of Sciences under grant no. 2018PM0017 and by the Strategic Priority Research Program of the Chinese Academy of Science ''Multi-waveband Gravitational Wave Universe'' (Grant No. XDB23040000).

%%%%%%%%%%%%%%%%%%%%%%%%%%%%%%%%%%%%%%%%%%%%%%%%%%

%%%%%%%%%%%%%%%%%%%% REFERENCES %%%%%%%%%%%%%%%%%%

% The best way to enter references is to use BibTeX:

\bibliographystyle{mnras}
\bibliography{bibliografia} % if your bibtex file is called example.bib

%%%%%%%%%%%%%%%%%%%%%%%%%%%%%%%%%%%%%%%%%%%%%%%%%%

%%%%%%%%%%%%%%%%% APPENDICES %%%%%%%%%%%%%%%%%%%%%

\appendix

%%%%%%%%%%%%%%%%%%%%%%%%%%%%%%%%%%%%%%%%%%%%%%%%%%

% Don't change these lines
\bsp	% typesetting comment
\label{lastpage}
\end{document}